\documentclass[prl,twocolumn,showpacs,preprintnumbers,amsmath,amssymb,floatfix]{revtex4}

\usepackage{subfigure,graphicx,epsfig,amsmath,amsfonts,amssymb,xcolor,slashed,ulem}

\newcommand{\be}{\begin{equation}}
\newcommand{\ee}{\end{equation}}
\newcommand{\ba}{\begin{eqnarray}}
\newcommand{\ea}{\end{eqnarray}}
\newcommand{\nn}{\nonumber}
\newcommand{\mev}{\textrm{ MeV}}
\newcommand{\gev}{\textrm{ GeV}}
\newcommand{\JP}{J/\psi}
\newcommand{\pa}{P_c(4380)}
\newcommand{\pb}{P_c(4450)}
\newcommand{\decay}{\Lambda_b\to J/\psi\,K^- p}

\begin{document}

\title{ On the hidden charm pentaquark-s in $\Lambda_b \to J/\psi K^- p$ decay}
 
\author{L. Roca}
\affiliation{Departamento de F\'isica, Universidad de Murcia, E-30100 Murcia, Spain}

\author{E.~Oset}

\affiliation{Departamento de
F\'{\i}sica Te\'orica and IFIC, Centro Mixto Universidad de
Valencia-CSIC Institutos de Investigaci\'on de Paterna, Aptdo.
22085, 46071 Valencia, Spain}

\date{\today}

\begin{abstract}

In a previous work we presented 
 a theoretical analysis of the $\Lambda_b \to J/\psi K^- p$ reaction from where a recent experiment by the LHCb collaboration at CERN claimed the existence of 
two hidden charm pentaquarks, $P_c(4380)^+$ and 
$P_c(4450)^+$. 
In that work we focused only on the $\Lambda(1405)$ and $P_c(4450)^+$ signals and discussed the possible explanation of this pentaquark state 
within the picture of a dynamical meson-baryon molecule made up mostly from   $\bar D^* \Sigma_c$ and $\bar D^* \Sigma^*_c$ components.
In the present work we improve upon the previous one by considering the total  $K^- p$ and $J/\psi  p$ data including all the relevant resonances contributing to the spectra, and discuss the possible nature of both $P_c(4380)^+$ and 
$P_c(4450)^+$. We also discuss several important topics, like the effect of the contact term in the reaction, the viability of reproducing the data without the  $P_c(4380)^+$ and the possible quantum numbers assignment to these pentaquarks.

\end{abstract}

\maketitle


\section{Introduction}

In the early beginning of the quark model Gell-Mann and Zweig already mentioned that apart from the standard $q \bar q$ and $q q q $ configurations for mesons and baryons respectively, there could exist some multiquark configurations \cite{GellMann:1964nj,zweig}. Concrete calculations for the case of pentaquarks were done by Hogaasen and Sorba \cite{Hogaasen:1978jw} and Strotmann \cite{Strottman:1979qu}. Much excitement was generated by the claims of an observation of a 
pentaquark state, $\Theta^+ $, at Spring8/Osaka in a photonuclear reaction \cite{Nakano:2003qx}. A similar method was used to analyze the $\gamma d \to p n K^+ K^- $ reaction where again a claim was made for the $\Theta^+ $ pentaquark \cite{Nakano:2008ee}. After a period of excitement where the peak was observed in most laboratories, searches with better statistics and analysis methods started to report negative results and the issue was closed. A detailed report on this issue can be seen in \cite{Hicks:2012zz}. An important work to clarify the issue was done in  \cite{Torres:2010jh,MartinezTorres:2010zzb}, were it was shown that the experimental peak observed in \cite{Nakano:2008ee} was a consequence of the analysis method in \cite{Nakano:2008ee}, where only the $K^+$ and $ K^- $ were observed and the $K^+ n$ invariant mass was constructed with a prescription for the unmeasured $n$ momentum, that was incorrect and artificially produced a peak in the ``$\Theta^+ $" region.

   With this precedent, the claim of two pentaquark states in the LHCb experiment \cite{exp,Aaij:2015fea} should have been taken with caution, but given the thoroughness of the experimental analysis, the result was difficult to challenge. Yet, issues concerning the implementation of unitarity in the experimental analysis, and the lack of a tree level contribution, which is unavoidable from a theoretical point of view, were raised \cite{rio}. One of the aims of the present paper is to discuss in detail these issues and show how some accidental circumstances make the present experimental analysis overcome this problem, thus providing extra support to the experimental claims. 

     The precedent of the former ``$\Theta^+ $" pentaquark unjustified claims has not prevented a wave of excitement among theoreticians, who have proposed a variety of possible explanations for these two states. One of the reasons for it is that predictions of hidden charm baryon states had been made before.  Indeed, in 
\cite{Wu:2010vk,Wu:2010jy} baryon states of hidden charm were found in the study of the interaction of the $\bar{D}\Sigma_{c}$-$\bar D\Lambda_c$,
$\bar{D}^{*}\Sigma_{c}$-$\bar D^*\Lambda_c$ coupled channels as the main building blocks, together with the $\eta_c N$ and $J/\psi N$ states, plus decay channels in the light sector. Related studies were done in \cite{Yang:2011wz}, where bound states of $\bar{D}\Sigma_c$ and $\bar{D}^*\Sigma_c$ were also found. In \cite{Garcia-Recio:2013gaa}, using an admixture of SU(6) and Heavy Quark Spin symmetry, HQSS, states of hidden charm similar to those predicted in \cite{Wu:2010vk,Wu:2010jy} were also found. Further studies were done and in ~\cite{hiddencharm} similar results to those of \cite{Wu:2010vk,Wu:2010jy} were found,  using HQSS and the local hidden gauge approach as tools to evaluate the matrix elements of the interaction. A quark model was used in \cite{Yuan:2012wz}, where some hidden charm baryons states were also obtained. All these works share similar qualitative results, but differ in the predictions of the masses of the particles by as much as 200 MeV up in \cite{Yuan:2012wz} to 200 MeV down in  \cite{Garcia-Recio:2013gaa} with respect to those found in \cite{Wu:2010vk,Wu:2010jy,hiddencharm}. Further work on this line is done in \cite{uchinohc}, using an admixture of Vector-Baryon and Pseudoscalar-Baryon states in coupled channels which allow one to have a better hold on the decay width of the states.

After the experimental papers, theoretical papers on the issue have proliferated, proposing many options.  Meson-baryon molecules have been suggested in \cite{Chen:2015loa,Roca:2015dva,He:2015cea,Huang:2015uda,Meissner:2015mza,Xiao:2015fia,Chen:2015moa,
Eides:2015dtr,Yang:2015bmv}. Pentaquark states of diquark-diquark-antiquark nature have been suggested in ~\cite{Maiani:2015vwa,Ghosh:2015ksa,Anisovich:2015cia,Wang:2015epa,Wang:2015ixb}, compact diquark-triquark pentaquarks in ~\cite{Lebed:2015tna,Zhu:2015bba},  $\bar{D}$-soliton states in \cite{Scoccola:2015nia}, genuine multiquark states in ~\cite{Mironov:2015ica,Gerasyuta:2015djk},
and some papers have suggested that the peaks could be related to kinematic effects of a triangle singularity~\cite{Guo:2015umn,Liu:2015fea,Mikhasenko:2015vca}. Summaries of the theoretical and experimental work done can be seen in Refs.~\cite{Burns:2015dwa,Wu:2015nhv,Stone:2015iba,Chen:2016heh,osetnew}, and in particular in the thorough review on the subject  \cite{Chen:2016qju}.

\section{Formalism}
\label{sec:formalism}

The core of the present analysis is the identification of the most relevant mechanisms contributing to the amplitude to describe the  $\Lambda_b \to J/\psi K^- p$
experimental data. We do not intend to obtain a better fit than the one carried out in the experimental analysis \cite{exp}, which included many different $\Lambda$ resonances in addition to the pentaquarks, considering possible different quantum number assignments for them and all angular dependences relative to the decay products of the $\JP$.
Our intention is to implement a good enough amplitude, but as simple as possible in order to discuss 
the relevant theoretical issues regarding the role played by the different contributions to the  $\Lambda_b \to J/\psi K^- p$, specially the pentaquarks and the dominant $\Lambda$ resonances.
Nevertheless, regarding the $K^- p$ and $\JP\,p$  invariant mass distributions our approach is quite accurate since the  angular information mentioned above, considered by the experimental analysis, is integrated in these observables.
The procedure followed in the present work is also rather different from the one followed in \cite{exp}. Instead of using the helicity formalism, we construct explicit amplitudes suited to the excitation of the different resonances and their quantum numbers. In addition, we explicitly consider the $K^- p$ formation at the tree level and its interaction with coupled channels in s-wave.

\begin{figure}[tbp]
     \centering
     \subfigure[]{
          \label{fig:diag1a}
          \includegraphics[width=.7\linewidth]{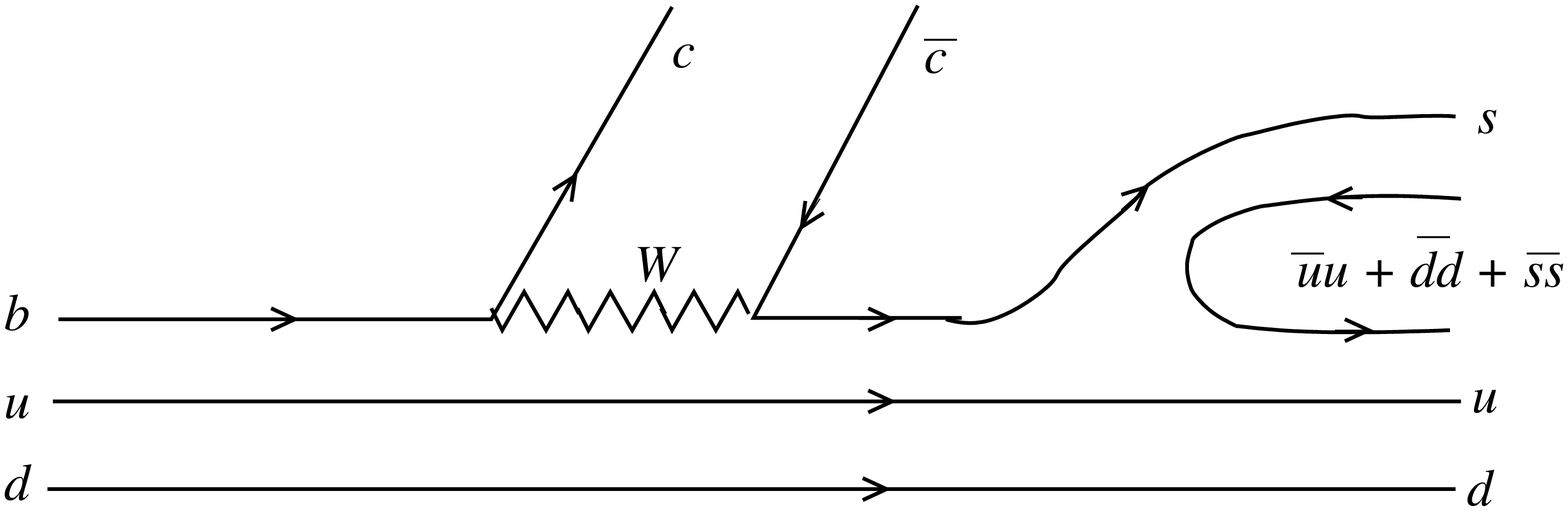}}\\
     \subfigure[]{
          \label{fig:diag1b}
          \includegraphics[width=.45\linewidth]{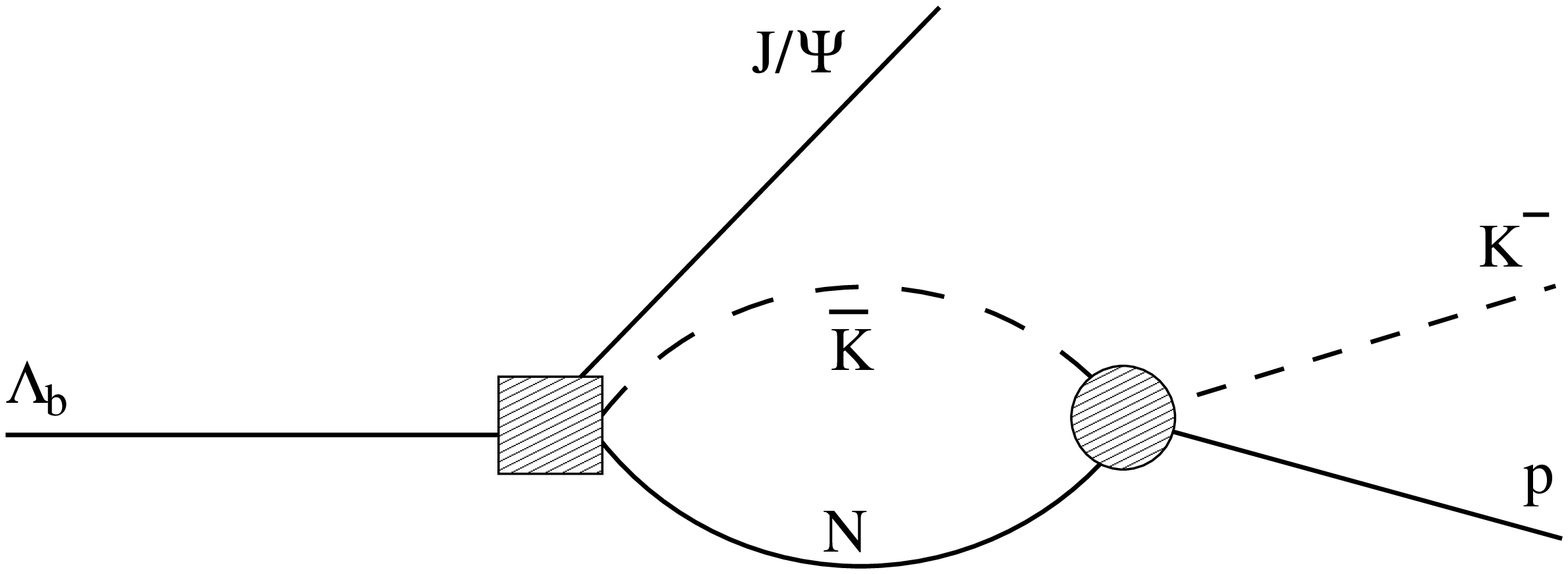}}
      \subfigure[]{
          \label{fig:diag1c}
          \includegraphics[width=.45\linewidth]{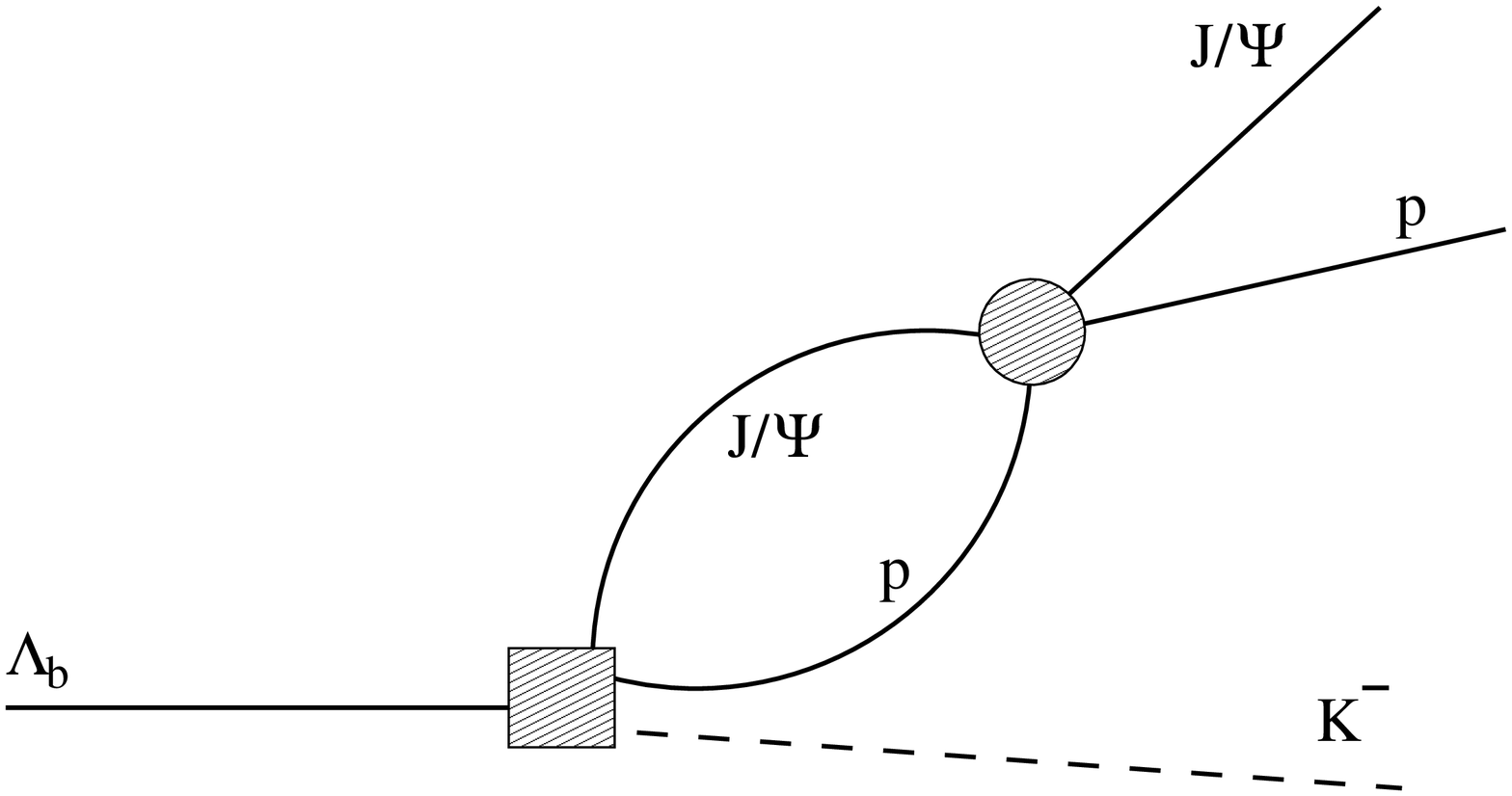}}
    \caption{Mechanisms for the $\Lambda_b\to J/\psi K^-p$ reaction implementing the final state interaction}\label{fig:diags}
\end{figure}

In ref.~\cite{Roca:2015dva} only the dynamically generated resonances, $\Lambda(1405)$ and $P_c(4450)$, were included,
and it was shown in \cite{rocamai} that the relevant mechanisms in the 
$\decay$ decay 
are those depicted in 
Figs.~\ref{fig:diag1a} and \ref{fig:diag1b}.
The $K^-p$ is produced at the quark level as depicted in  Fig.~\ref{fig:diag1a}.  
It is also interesting to recall that in \cite{rocamai} it was also found that the $\JP  \pi \Sigma$ states were not produced at this tree level.
First the weak interaction produces the $c \bar c$ state ($J/\psi$)  and an $s$ quark and then
 the $u$, $d$ and $s$ remaining quarks hadronize into a meson-baryon pair. Note that the $u$ and $d$ quarks are spectators during
the entire process, as justified in
 ref.~\cite{rocamai}.
Within the framework of the chiral unitary approach, the 
  $\Lambda(1405)$ resonance, of crucial role in the present decay, is  produced dynamically from the final $K^- p$ state interaction implementing unitarity in the different meson baryon channels with the $\Lambda(1405)$ quantum numbers.
In this way, and without the need to include the $\Lambda(1405)$ resonances   as explicit degrees of freedom, two poles were obtained 
for the $\Lambda(1405)$ resonance \cite{Jido:2003cb}.
Recently the positions of the poles were more finely obtained
at $\sqrt{s}=1352-48i$~MeV 
and $1419-29i$~MeV \cite{Roca:2013cca}, with the lowest mass pole coupling mostly to $\pi\Sigma$ and the highest mass one
to $\bar K N$. Therefore its implementation into the $\decay$ decay is depicted in Fig.~\ref{fig:diag1b}, and it is given by
\begin{align}\label{eqn:tgt1405}
T^{\Lambda(1405)}(M_{K^-p})=\alpha_1 G_{K^-p}(M_{K^-p})
\,t^{I=0}_{\bar K N,\bar K N}(M_{K^-p}) \,,
\end{align}
where $M_{K^-p}$ ($M_{\JP p}$) stands for the $K^-p$ ($\JP p$) invariant mass, $G_{K^-p}$ is the $K^-p$ loop function and 
$t^{I=0}_{\bar K N,\bar K N}(M_{K^-p})$ the s-wave, isospin 0, $\bar K$-nucleon unitarized  scattering amplitude from ref.~\cite{Roca:2013cca}, (thick circle in Fig.~\ref{fig:diag1b}).
The parameter $\alpha_1$ is a free parameter, to be fitted later on, accounting for the elementary production process, Fig.~\ref{fig:diag1a}.
Note that in addition to the previous amplitude we have to add to the total amplitude the $J/\psi\,K^-p$ tree level contact term contribution, Fig.~\ref{fig:diag1a}, which is shown in ref.~\cite{rocamai} to have the same weight, $\alpha_1$,  as the term in Eq.~(\ref{eqn:tgt1405}). This is a non-trivial result. 
Thus the contribution of the mechanisms in Fig.~\ref{fig:diag1a} and \ref{fig:diag1b} is
\be
\alpha_1 \left(1+G_{K^-p}(M_{K^-p})\,t^{I=0}_{\bar K N,\bar K N}(M_{K^-p})\right)
\label{eq:1plusGT}
\ee
Note that this 
tree level contribution, accounted for by the 1 addend in Eq.~(\ref{eq:1plusGT}), interferes with the amplitude of Eq.~(\ref{eqn:tgt1405}). 
Although constant nonresonant terms were used in the fit of 
\cite{exp}, the output of the fit did not return a significant contribution of these terms. We shall try to understand this feature, and this issue will be widely discussed in the results section.

As mentioned in the Introduction, in refs.~\cite{hiddencharm,Wu:2010vk,uchinohc},
several poles were obtained (see table~II of ref.~\cite{hiddencharm}
and table~8 of ref.~\cite{uchinohc}) when implementing unitarity in coupled channels in s-wave considering the channels 
$\JP N$, $\eta_c N$,
 $\bar D B$ and $\bar D^* B$, with
$B$ baryon charmed states belonging to the 20
representations of $SU(4)$ with $J^P=1/2^-$ and $3/2^-$.
Exploring the possibility that some of these poles (or a mixture of several ones) could correspond to the experimentally found $P_c$ pentaquarks is one of the aims of the present work. In such a  case, the production of the pentaquark in our model would proceed through the mechanism depicted in 
Fig.~\ref{fig:diag1c}:  The $\JP\,p$ pair initially produced  
undergoes final state interaction, which is accounted for by the 
 $\JP\,p \to\JP\, p$ unitarized scattering amplitude represented by the thick circular dot in Fig.~\ref{fig:diag1c}. 
 In this particular case, the $\JP\,p \to\JP\, p$ unitarized scattering amplitude resembles very much a Breit-Wigner \cite{hiddencharm}, therefore, and for the numerical evaluation carried out in the present work, the mechanism of Fig.~\ref{fig:diag1c} can be effectively
 accounted for by a term proportional to

\noindent
\be
 \alpha_i\,G_{\JP p}\,\frac{g^2_{\JP\,p}}{M_{\JP\,p}-m_{P_c}+i \frac{\Gamma_{P_c}}{2}}
\label{eq:tJpsi}
\ee
with $G_{\JP p}$ the $\JP p$ loop function and $m_{P_c}$ ($\Gamma_{P_C}$) the mass (width) of either of the pentaquarks.
The term in Eq.~(\ref{eq:tJpsi}) has to be multiplied by a momentum structure which depends on the different possible quantum number assignment of the pentaquarks. (See Appendix~\ref{appendix:interf} for explicit details).
 The pole positions of the amplitudes obtained in 
 refs.~\cite{hiddencharm,Wu:2010vk,uchinohc} 
 provide directly  $m_{P_c}$ and $\Gamma_{P_c}$, but with uncertainties in the mass of the order of $200\mev$ \cite{uchinohc}. Therefore we fine tune these values to the  experimental results of ref.~\cite{exp}, $m_{P_c}=4380\mev$ and $\Gamma_{P_c}=205\mev$ for the lowest pentaquark and $m_{P_c}=4449.8\mev$ and $\Gamma_{P_c}=40\mev$ for the highest one, which lie indeed in between the different values obtained in refs.~\cite{hiddencharm,Wu:2010vk,uchinohc}.
The coupling of the dynamically generated resonance to $\JP\,p$, $g_{\JP\,p}$ in Eq.~\eqref{eq:tJpsi}, was determined to be of the order of 0.5 in  refs.~\cite{hiddencharm,Wu:2010vk,Roca:2015dva}. 
However, we have included in the amplitude of Eq.~(\ref{eq:tJpsi}) a free parameter $\alpha_i$, ($i=2$ for $P_c(4450)$ and $i=3$ for $P_c(4380)$), to better fit the experimental data, which should have a natural value of the order of 1.
The presence of the  $\JP p$ loop function,
$G_{\JP p}$, in Eq.~(\ref{eq:tJpsi}) has to be included when
assuming the resonances to be dynamically generated, since in such a case it is always produced by the scattering of an initial  $\JP p$ pair. Nevertheless, the $G_{\JP p}$ factor has little impact in the global fit.

We can also consider the scenario where one or both pentaquarks 
have $J^P$ different to $1/2^-$ or $3/2^-$ which are the quantum numbers of the states generated by the chiral unitary approach \cite{hiddencharm,Wu:2010vk,uchinohc}.
 In this scenario, and if the $\JP p$ pair was in p-wave, the corresponding pentaquark would carry $J^P=1/2^+$, $3/2^+$ or $5/2^+$ and its contribution should be added to the previous terms leading to the $\Lambda(1405)$ (see  Appendix). Nonetheless, and as we will explain below, since the qualitative output of our fits are similar irrespective of the quantum number assignments for the pentaquarks, we will only work out the $5/2^+$ case (in addition to the previous $1/2^-$ and $3/2^-$)     for simplicity since only one possible partial wave for the kaon is possible, ($L'=2)$, in that case, (see  Appendix~\ref{appendix:interf}, table~~\ref{tab:appendix1}, for details).
\begin{figure}[tbp]
     \centering
     \subfigure[]{
          \label{fig:diagPLa}
          \includegraphics[width=.45\linewidth]{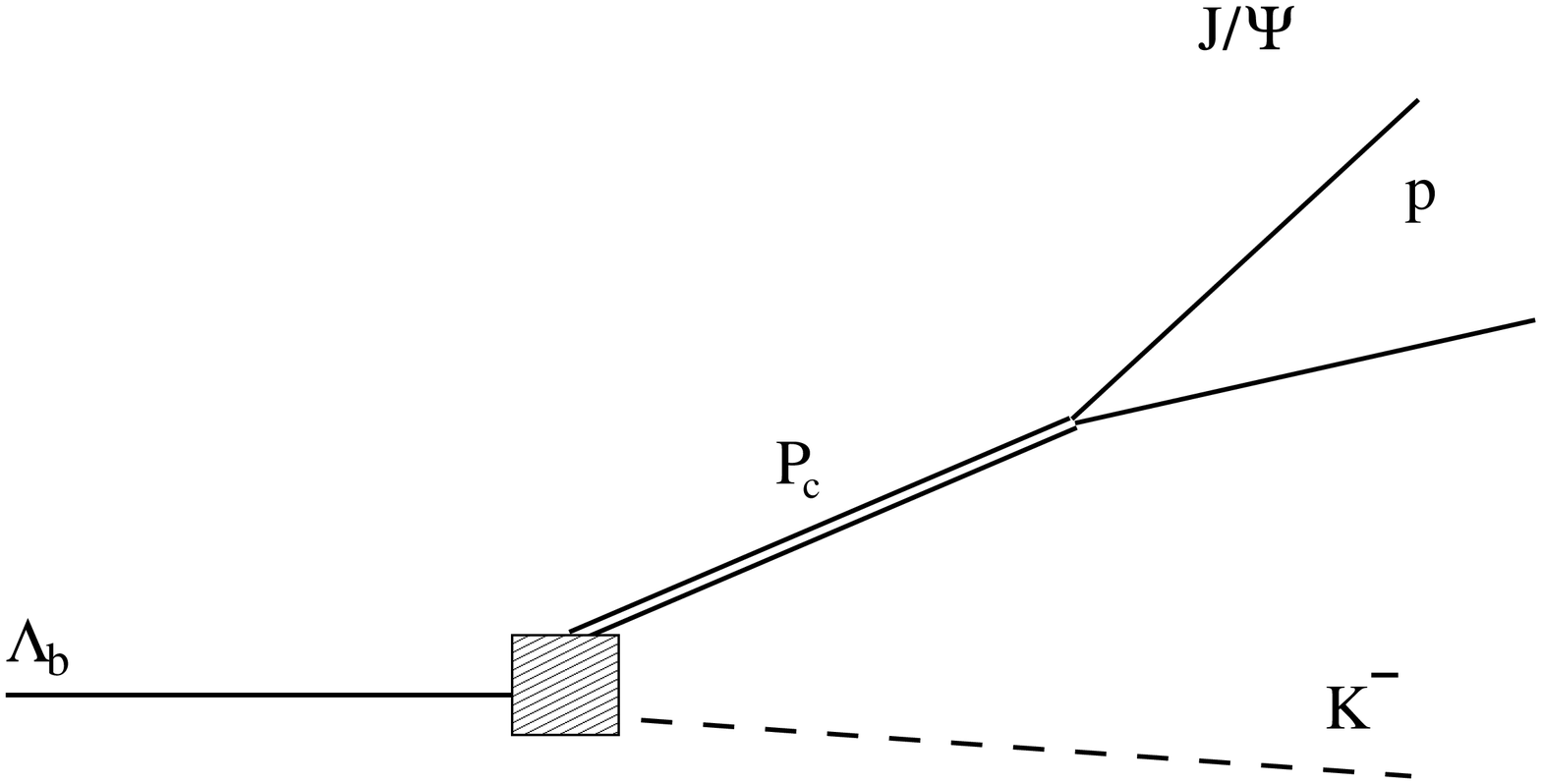}}
     \subfigure[]{
          \label{fig:diagPb}
          \includegraphics[width=.45\linewidth]{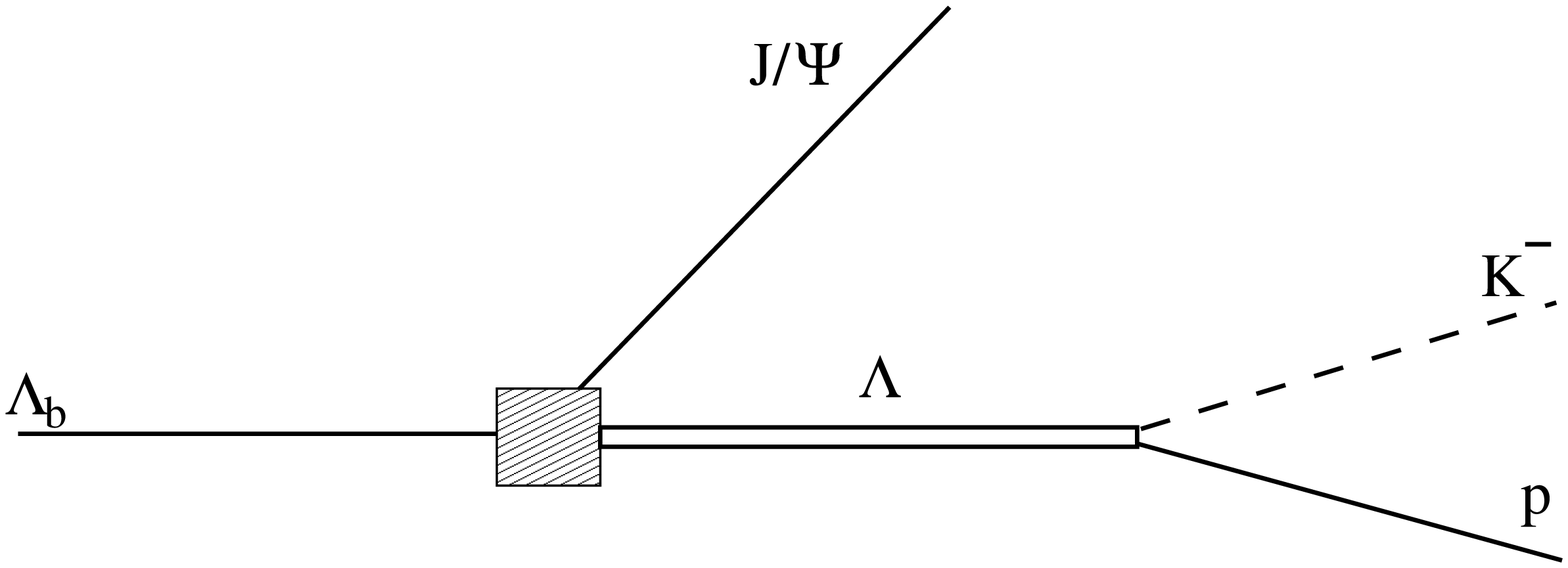}}
    \caption{Mechanisms for the $\Lambda_b\to J/\psi K^-p$ reaction considering explicit, non dynamical, pentaquarks (a) and $\Lambda$ resonances (b).}\label{fig:diagsPL}
\end{figure}
 In this case the mechanism for the pentaquark production would be simply that depicted in Fig.~\ref{fig:diagsPL}a and, in order to establish its amplitude, we  must remove $G_{\JP p}$ in Eq.~\eqref{eq:tJpsi} and multiply it by a momentum dependence according to its quantum numbers as shown in Eqs.~(\ref{eq:Tstructure}), (\ref{eq:D52p}) and (\ref{eq:coefff}).

On the other hand, in the analysis of ref.~\cite{Roca:2015dva} no further $\Lambda$ resonances, (in addition to the $\Lambda(1405)$), were considered since that study was based only on the $\Lambda(1405)$ and the pentaquark filtered signal. However, since in the present work we aim at reproducing the total $K^- p$  and $\JP\,p$ invariant mass, we must add the relevant $\Lambda$ contributions to the process, Fig.~\ref{fig:diagsPL}b.
In ref.~\cite{exp}, 13 different $\Lambda$ resonances were included in the fit. In our analysis it is enough to consider only those which gave a sizable contribution to the final cross section in \cite{exp} which, in addition to the $\Lambda(1405)$ discussed above,
turn out to be $\Lambda(1520)$ ($3/2^-$), $\Lambda(1600)$ ($1/2^+$), $\Lambda(1690)$ ($3/2^-$), $\Lambda(1800)$ ($1/2^-$) and $\Lambda(1810)$ ($1/2^+$).

The $\Lambda$ resonances (except the $\Lambda(1405)$ which is explained above, see Eq.~(\ref{eqn:tgt1405})) are parametrized by a Breit-Wigner shape with 
Flatt\'e parametrization of the width by
\be
\label{eqn:BWLambda}
t_{\Lambda_i}=\alpha_i
\frac{1}{{M_{K^-p}-m_{\Lambda_i}+i\frac{\Gamma_{\Lambda_i}(M_{K^-p})}{2}}}
\ee
up to a factor containing the spin and momentum structure as explained in the Appendix.
 The widths of the $\Lambda$ resonances have been
   taken energy dependent in the following way:
\ba
\Gamma_{\Lambda_i}= \Gamma_{o_i} \frac{m_{\Lambda_i}}{M_{K^-p}}\sum_j f_j \left(\frac{q_j}{q_{o_j}}\right)^{2L_j+1}B(L_j,q_j,q_{o_j})^2
\label{eqn:widthenergy}
\ea
with $\Gamma_{o_i}$ the on shell width, and $f_i$ the branching ratio of the $\Lambda_i$ into the dominant decay channels, $j$, obtained from the PDG \cite{pdg}. In Eq.~(\ref{eqn:widthenergy}) $B(L,q,q_o)$ is the Blatt-Weisskopf penetration factor for L-wave \cite{blattweisskopf},
where $q$ ($q_o$) is the $K$ or $p$ momentum at the $K^-p$ center of mass frame at $M_{K^-p}$ ($m_i$).

Furthermore contact terms with the different spin and angular momentum of the $\JP$ and kaon are also considered as explained in the Appendix.
These contact terms were found to be negligible in the experimental fit \cite{exp} but we will explain in the results section that some of them could play indeed an important role.

Altogether, the $\Lambda_b \to J/\psi K^- p$ differential decay rate is given by

\begin{align}\label{eqn:dGammadM}
\frac{d^2\Gamma_{\Lambda_b \to J/\psi K^- p}(M_{K^-p},M_{\JP p})}
{dM_{K^-p} dM_{\JP p} }
=  \frac{1}{16\pi^3}\frac{m_p}{m_{\Lambda_b}^2}\,\times \nn\\
\times M_{K^-p}M_{\JP p}\, \left|T(M_{K^-p},M_{\JP p})\right|^2\,.
\end{align}
The interferences between the different $\Lambda$ mechanisms and the pentaquarks and contact terms depend on the different quantum numbers of the particular $\Lambda$ resonances and the pentaquarks. The total amplitude squared
$|T(M_{K^-p}M_{\JP p})|^2$ (of course averaged over initial spins and added over final ones) is given in the Appendix for the different cases.
The $\alpha_i$ and $C_i$ in Eqs.~(\ref{eq:coeffa}),
(\ref{eq:coeffb}), (\ref{eq:coeffc}), (\ref{eq:coeffe}) and (\ref{eq:coefff}) and  are free parameters to be 
fitted in our analysis of the following section.

\section{Results and discussion}

We have carried out different fits to the experimental data \cite{exp} 
considering the following possibilities for the spin-parity of the pentaquarks $(J^P_A,J^P_B)$ where $J^P_A$ and $J^P_B$ stand for the spin-parity of the $P_c(4380)$ and $P_c(4450)$ respectively:
$(1/2^-,1/2^-)$,  $(1/2^-,3/2^-)$, $(3/2^-,1/2^-)$, $(3/2^-,3/2^-)$, $(3/2^-,5/2^+)$, $(5/2^+,3/2^-)$, $(1/2^-,5/2^+)$, $(5/2^+,1/2^-)$ and $(5/2^+,5/2^+)$.
Nevertheless, and advancing some results, neither of them provides a remarkably better fit than the rest,
and all of them produce qualitatively similar results.
However, experimentally \cite{exp} a best fit was obtained for $J^P$ assignments of  $(3/2^-,5/2^+)$, and also acceptable fits were obtained for $(3/2^+,5/2^-)$  and $(5/2^+,3/2^-)$ and ``other combinations are less likely'', but there was a clear preference for two states with opposite parity \cite{exp}.
The experimental analysis \cite{exp} is more complete than the one carried out in the present work since it takes into consideration more $\Lambda$ resonances and, specially, all angular dependences relative to the decay products of the $\JP$, which we do not consider. 
However the angular dependence is not relevant for the invariant mass distributions, which are the observables we fit, therefore the extra conclusions drawn by the experimental analysis \cite{exp} about the 
$J^P$ of the pentaquarks must certainly come from that extra information used in the experimental analysis, (note that there are more than 150 parameters fitted in \cite{exp} while there are just 19 in our analysis).
Yet, it is worth stressing again that it is not the intent of the present work to improve or even be on 
a par with the already good experimental analysis. Our focus is on specific theoretical issues for which  the approach followed here is good enough.
Therefore, since the qualitative discussion is similar irrespectively of the $J^P$ pentaquark assignments in our model, we will discuss upon results for $(3/2^-,3/2^-)$ assignment, one of the chiral unitary options, unless other case is explicitly stated.


\begin{figure*}[t!]
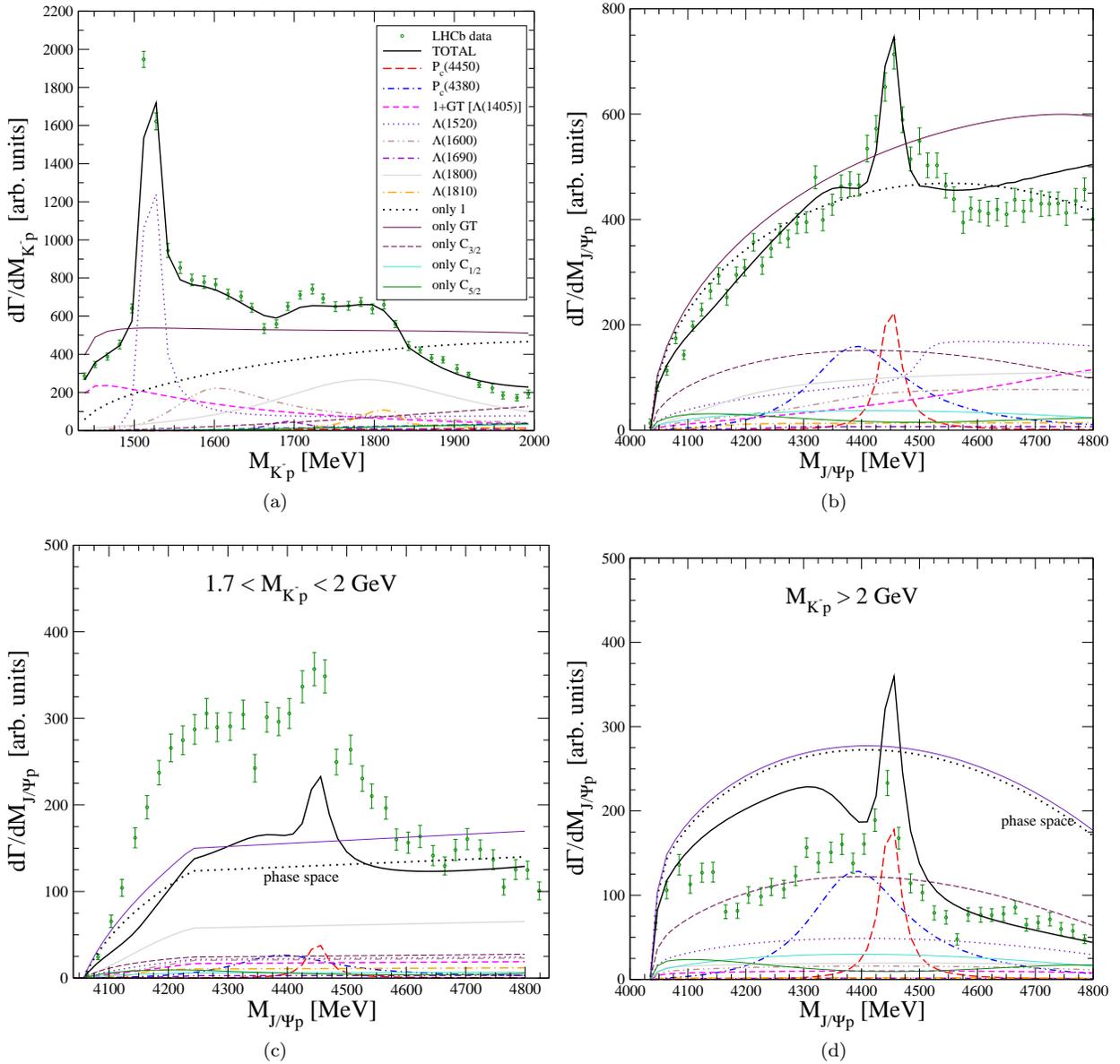

     \centering
     \subfigure[]{
          \label{fig:res1a}
          \includegraphics[width=.45\linewidth]{figRUN104_a.eps}}
     \subfigure[]{
          \label{fig:res1b}
          \includegraphics[width=.45\linewidth]{figRUN104_b.eps}}
\\      \subfigure[]{
          \label{fig:res1c}
          \includegraphics[width=.45\linewidth]{figRUN104_c.eps}}
     \subfigure[]{
          \label{fig:res1d}
          \includegraphics[width=.45\linewidth]{figRUN104_d.eps}}
    \caption{(Color online). Panels (a) and (b): experimental data used to fit the model and results from the fit with the full model for the $K^- p$ and $\JP\,p$  invariant mass distributions respectively and for  the $(3/2^-,3/2^-)$ case . Panels (c) and (d): 
    Predictions for the $\JP\,p$ distribution implementing the kinematic cuts shown in the figures. (The data from panels (c) and (d) are not fitted).}
       \label{fig:res1}
\end{figure*}

First we show in the (a) and (b) panels of Fig.~\ref{fig:res1} the result of the fit to the experimental data \cite{exp} and the individual contributions of the different resonances. (The $1+GT$ label stands for the term $1+G_{K^-p}(M_{K^-p})\,t^{I=0}_{\bar K N,\bar K N}(M_{K^-p})$ in 
Eqs.~(\ref{eq:1plusGT}) and (\ref{eq:coeffa}) which essentially produces the $\Lambda(1405)$, as explained in the previous section).
We can see that the global fit is quite fair, given the simplified version of the model compared to the analysis done in the experimental work \cite{exp}. Note specially the important strength of both pentaquarks, $P_c(4380)$ and $P_c(4450)$ in the $\JP p$ mass distribution. We have only fitted the $K^- p$ mass distributions up to $M_{K^-p}<2\gev$ and the $\JP p$ up to $4.8\gev$ since our chiral unitary model for the $\Lambda(1405)$ cannot be extrapolated to further higher energies and the reduced range is preferable and sufficient for our discussion.
Panels (c) and (d) in Fig.~\ref{fig:res1} reflect data and results 
for $\JP p$ mass distribution implementing the kinematic cuts
$1.70\gev<M_{K^-p}<2\gev$ and  $M_{K^-p}>2.0\gev$ respectively. (See figure 8 of ref.~\cite{exp}).  The data in panels (c) and (d) 
are not fitted, thus the curves therein are output of the calculation.
 It is worth noting that the experimental data  for the $K^- p$ and $\JP\,p$ invariant mass distributions are not corrected for experimental setup acceptance in ref.~\cite{exp}. However, in the experimental paper, phase-space curves
for Fig.~\ref{fig:res1}a and Fig.~\ref{fig:res1}b are provided, which of course are affected by the acceptance. Therefore, comparing those curves to the corresponding theoretical phase-space three body distribution, we have renormalized each experimental datum such that the phase-space agrees with the actual one and such that the areas below both invariant mass distributions are the same ({\it i.e.} same total $\Lambda_b$ width). 
This acceptance correction cannot be performed in the theoretical results in panels (c) and (d) since the acceptance modified phase-space is not provided for such kinematic cuts in \cite{exp}. This is one of the reasons of the rough agreement between our theoretical calculation and experiment in panels (c) and (d).
However, the main reason is that those cuts filter events in the higher part of the $K^- p$ spectrum which are little relevant in the global fit, or actually are not fitted at all.
 Nevertheless, it is worth noting that our results of Fig.~\ref{fig:res1}c and d will be used just qualitatively in the coming discussions below.

\begin{figure*}[t!]
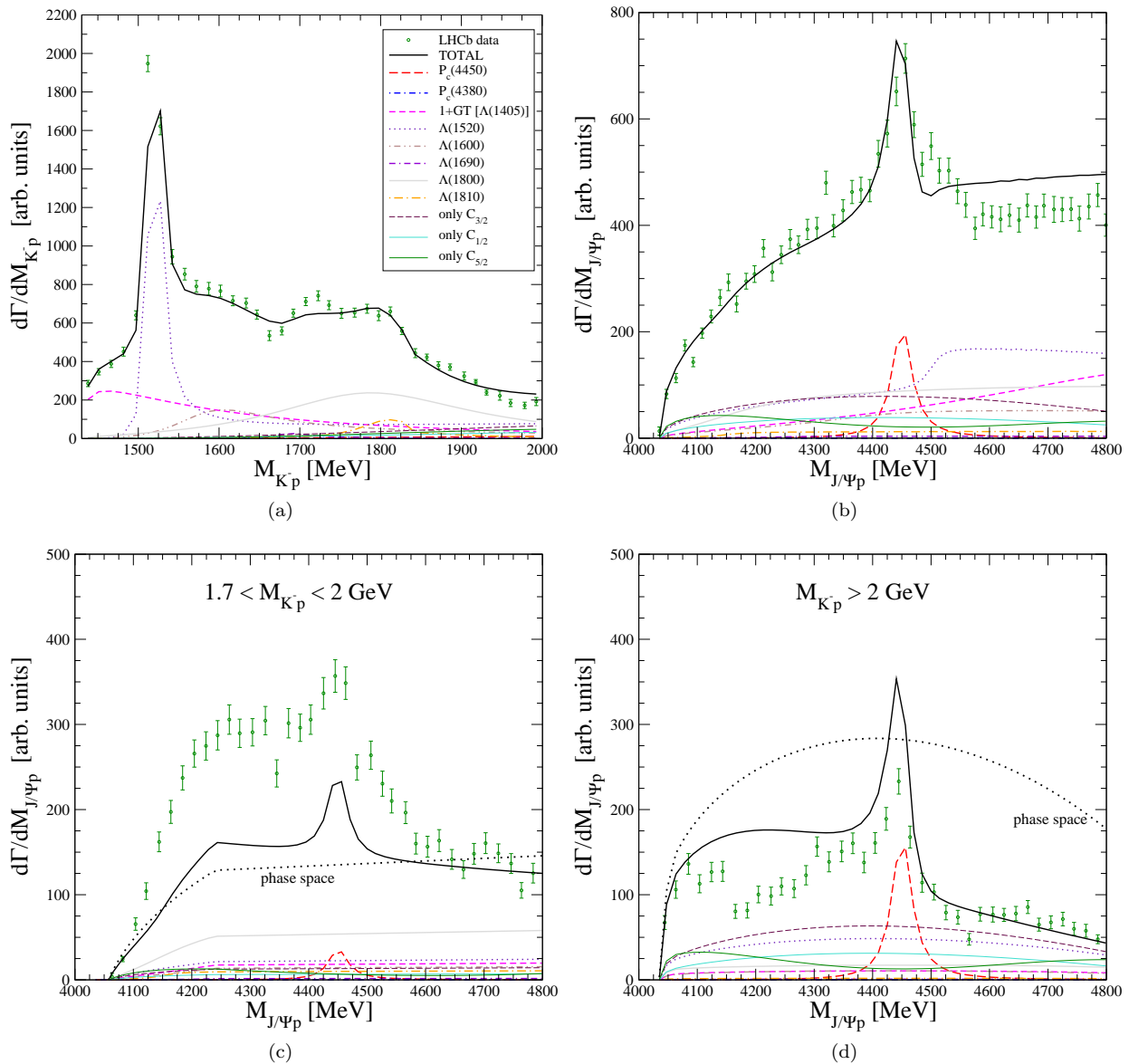

     \centering
     \subfigure[]{
          \label{fig:res2a}
          \includegraphics[width=.45\linewidth]{figRUN1044_a.eps}}
     \subfigure[]{
          \label{fig:res2b}
          \includegraphics[width=.45\linewidth]{figRUN1044_b.eps}}
\\      \subfigure[]{
          \label{fig:res2c}
          \includegraphics[width=.45\linewidth]{figRUN1044_c.eps}}
     \subfigure[]{
          \label{fig:res2d}
          \includegraphics[width=.45\linewidth]{figRUN1044_d.eps}}
    \caption{(Color online). Same as Fig.~\ref{fig:res1} but removing the $\pa$ from the fit.}
       \label{fig:res2}
\end{figure*}

In ref.~\cite{Roca:2015dva} it was pointed out  that the experimental support for the existence of the $P_c(4380)$ state was not as clear as 
for the $P_c(4450)$ one. The reason was that the  Argand plot for the $P_c(4380)$ is not as clean
 as that of the narrow $P_c(4450)$ \cite{exp}. On the other hand there is not a clear bump for the $P_c(4380)$ in the total $\JP p$ experimental invariant mass distribution. To shed some light into this issue we next carry out a fit removing the $P_c(4380)$ term. The result is shown in Fig.~\ref{fig:res2}a and b.  This fit without $P_c(4380)$ is just slightly worse compared to that in Figs.~\ref{fig:res1}a and b, (about a 20\% bigger $\chi^2/\textrm{dof}$), but it 
  is specially very similar or even better in the lowest region of the $\JP p$ mass distribution. 
  We have traced the reason for the good agreement in the low  
$\JP p$ mass region when removing the $P_c(4380)$ to the $J^P=5/2^+$ contact term (the 1 addend in Eq.~(\ref{eq:coefff})). (The shape of this contact term by itself corresponds to the label "only $C_{5/2}$" in the figures.) We see that with a slight increase of this contact term when the $P_c(4380)$ contact term is removed (compare curves "only $C_{5/2}$" between figures Figs.~\ref{fig:res1}b and Fig.~\ref{fig:res2}b), a similar effect than the one of the $P_c(4380)$ can be mostly accommodated.
Very similar results and conclusions are obtained for the other  $(J^P_A,J^P_B)$ possibilities studied in the present work.

   Therefore we must conclude that the fit to only the invariant mass distributions is not enough to draw firm conclusions about the existence of the $P_c(4380)$ state or the spin-parity of the pentaquark states, since many different $(J^P_A,J^P_B)$ possibilities yield similar results.
  The reason why the experimental analysis get  $(3/2^-,5/2^+)$ as the best option and $(3/2^+,5/2^-)$  and $(5/2^+,3/2^-)$ also acceptable 
 and ``other combinations are less likely'' \cite{exp} must then be traced to some other observable beyond just the   $K^- p$ and $\JP p$ mass distribution, like the angular dependence of the $\JP$ decay products, etc.

 On the other hand, if we next direct the attention to the phase space distribution in Fig.~\ref{fig:res2}c, we see that the kinematic cut implemented in that plot causes by itself a sharp kink around $M_{\JP p}\sim 4.2\gev$
which is responsible for  some of the accumulation of strength seen in the experimental data around
 that region. Therefore this apparent accumulation of strength at low $M_{\JP p}$ is not an indication by itself of the existence of a nearby resonance, like the $P_c(4380)$.

\begin{figure*}[t!]
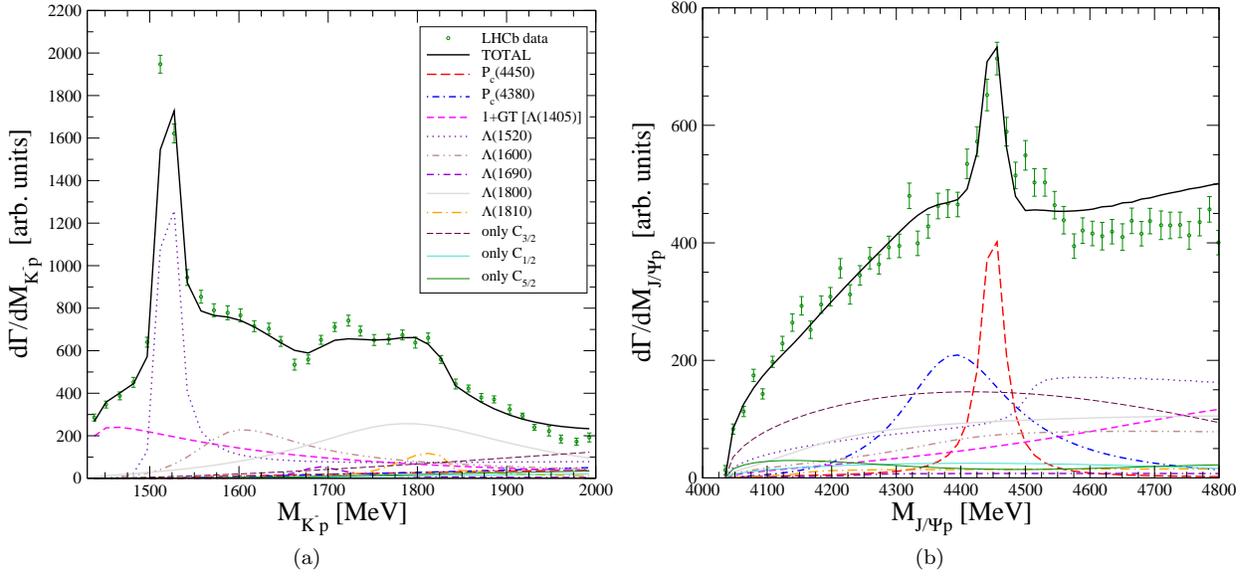

     \centering
     \subfigure[]{
          \label{fig:res3a}
          \includegraphics[width=.45\linewidth]{figRUN5_a.eps}}
     \subfigure[]{
          \label{fig:res3b}
          \includegraphics[width=.45\linewidth]{figRUN5_b.eps}}
    \caption{(Color online). Same as Fig.~\ref{fig:res1}a and b but with the $(3/2^-,5/2^+)$ configuration for the spin-parity of the pentaquarks}
       \label{fig:res3}
\end{figure*}

 In order to illustrate the similarity of the fits among different $J^P$ pentaquark assignments mentioned at the beginning of this section, we show in Fig.~\ref{fig:res3} 
 the result of the fit for the $(3/2^-,5/2^+)$ case which is the case for the best fit in the experimental work \cite{exp}.
In our case, however, we get  a fit with a very similar $\chi^2/\textrm{dof}$ to that in Fig.~\ref{fig:res1}a and b.

Let us next discuss one of the most important issues to be addressed in the present work, that is the effect of the contact term contribution of Fig.~\ref{fig:diag1a}, represented by the 1 addend in Eqs.~(\ref{eq:1plusGT}) and (\ref{eq:coeffa}).
Although a nonresonant term of this type was considered in the fit of
\cite{exp}, the output returned a negligible contribution from this source,
and, in principle, it could play an important role in the results. 
In the experimental fit \cite{exp},
 a Breit-Wigner shape is considered for the different resonances but
 for the $\Lambda(1405)$ the width has two components to account for the
 Flatt\'e effect.
  However, as explained above in the formalism section, in the chiral unitary approach this resonance is generated (actually two of them) from the  $K^- p$ interaction and accounted for by the 
$1+G_{K^-p}(M_{K^-p})\,t^{I=0}_{\bar K N,\bar K N}(M_{K^-p})$ in 
Eqs.~(\ref{eq:1plusGT}) and (\ref{eq:coeffa}), (we will just call it $1+GT$ term in the following).
Note that we include the 1 addend for the contact term which seems to be negligible in the final fit of \cite{exp}. 
Indeed, we see in Fig.~\ref{fig:res1}a and b that the contribution of the 1 addend by itself (curves labeled ``only 1'') is very large and it is crucial to produce the final $\Lambda(1405)$ strength and shape in the invariant mass distributions from its interference with the GT term
(curves labeled ``only GT''), which is also very large by itself and with a shape very different from a $\Lambda(1405)$ resonance.
However, we are going to show that a fortuitous combination of facts renders the approach of \cite{exp} very similar to the fully unitary approach that we follow here. 

Elaborating on this latter issue, let us recall that the two $\Lambda(1405)$ are basically obtained from the interaction of the coupled channels $\pi\Sigma$ and $\bar K N$. We shall call $T_{ij}$ the transition matrices from channels 1 ($\pi\Sigma$) and 2 ($\bar K N$).
Actually what one would expect to be approximately a Breit-Wigner is $T_{11}$ itself, not $(1+GT)_{11}$.
Therefore, in order to mimic the experimental approach to the $\Lambda(1405)$ we have performed a different fit implementing the substitution  
\begin{align}\label{eqn:TeqBW}
1+G_{K^-p}(M_{K^-p})\,&t^{I=0}_{\bar K N,\bar K N}(M_{K^-p}) \nn\\
&\longrightarrow 
\frac{1}{ {M_{K^-p}-m_R+i\frac{\Gamma_R}{2}}}
\end{align} 
with 
\begin{align}\label{eqn:GBW}
\Gamma_R=\Gamma_o\left(\frac{p_{\pi\Sigma}}
{p_{{{\pi\Sigma}}|_o}}\right)\frac{m_R}{M_{K^-p}}
+\alpha\, p_K\,\Theta(M_{K^-p}-m_K-m_p)
\end{align} 
where $m_R$, $\Gamma_o$ and $\alpha$ are adjusted to approximately 
reproduce the $t^{I=0}_{\bar K N,\bar K N}$ amplitude in the
 $\Lambda(1405)$ resonance region and $\Theta$ is the step function.
  The term proportional to $\alpha$ in Eq.~(\ref{eqn:GBW}) is included
   in order to account for the Flatt\'e effect, which is also
    incorporated in the analysis of ref.~\cite{exp}.
\begin{figure*}[t!]
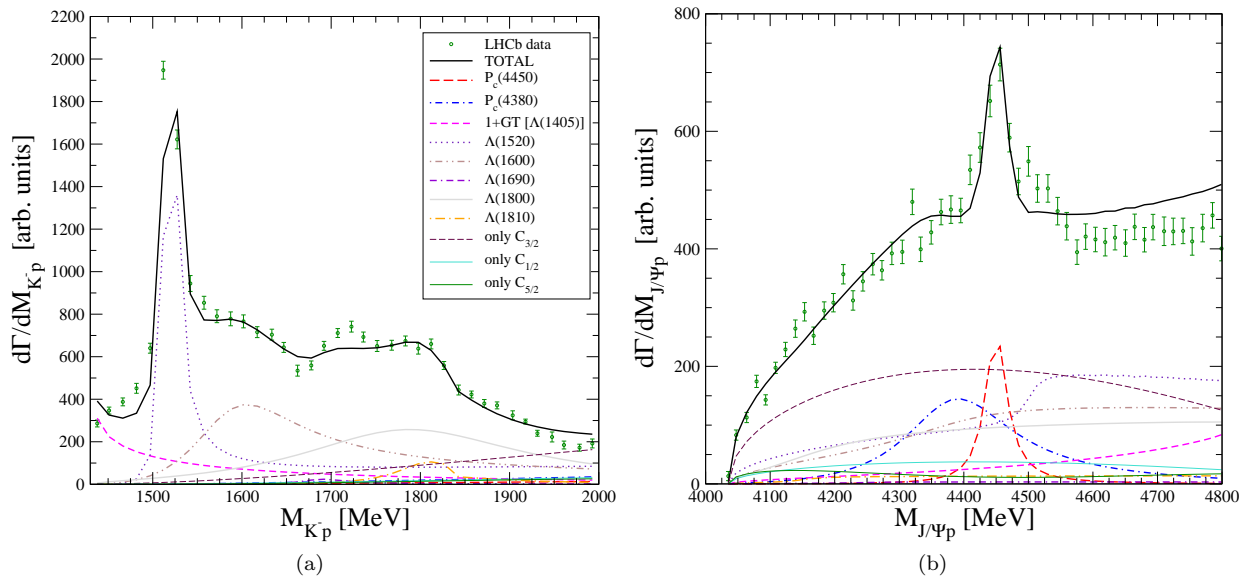

     \centering
     \subfigure[]{
          \label{fig:res4a}
          \includegraphics[width=.45\linewidth]{figRUN1045_a.eps}}
     \subfigure[]{
          \label{fig:res4b}
          \includegraphics[width=.45\linewidth]{figRUN1045_b.eps}}
    \caption{(Color online). Same as Fig.~\ref{fig:res1}a and b but
    carrying out the fit substituting the $1+GT$ term in
    Eqs.~(\ref{eq:1plusGT}) and (\ref{eq:coeffa}) by a Breit-Wigner $\Lambda(1405)$ amplitude
    with Flatt\'e parametrization of the width.}
       \label{fig:res4}
\end{figure*}
The result of the fit is shown in Fig.~\ref{fig:res4}. We can see that the result is very similar to Fig.~\ref{fig:res1}a,b, in spite of the fact that one is neglecting the contact term in the latter analysis.

\begin{figure}[t!]
     \centering
          \includegraphics[width=.8\linewidth]{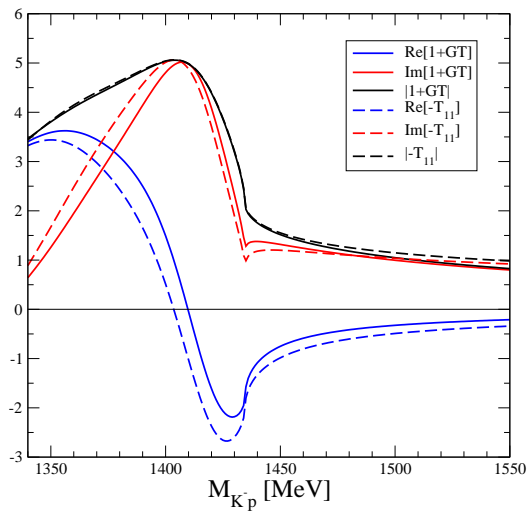}
    \caption{(Color online). Real part, imaginary part and absolute value of $1+G_{K^-p}\,t^{I=0}_{\bar K N,\bar K N}$ and $-t_{\bar K N,\bar K N}$, the later normalized such that the absolute value agrees in the region of the $\Lambda(1405)$.}
       \label{fig:unomasGT}
\end{figure}
This certainly requires a detailed explanation:
If we plot $(1+GT)_{11}$  and $-T_{11}$ we see in Fig.~\ref{fig:unomasGT} that they are very approximately proportional, (in that figure $-T_{11}$ has been multiplied by a constant factor such that the absolute values are the same at the $\Lambda(1405)$ mass). Actually the curves for the absolute value of  $(1+GT)_{11}$  and $-T_{11}$ are almost the same (up to  a global constant).
This is not true in a general case for a unitarized scattering  amplitude
since, in the chiral unitary approach and for two coupled channels,
we have
\begin{align}
&1+GT=V^{-1}T,\nn\\
&(1+GT)_{11}=(V^{-1}T)_{11}=\frac{T_{11}V_{22}-T_{12}V_{12}}{V_{11}V_{22}-V_{12}^2}.
\end{align}
However, it turns out that $V_{12}$ is about a factor 3 smaller than $V_{11}$ and $V_{22}$ and therefore, taking into account that $V_{11}$ is negative and smooth in the energy region considered, we have
\begin{align}\label{eqn:1GTmT}
&(1+GT)_{11}\simeq\frac{T_{11}}{V_{11}}\propto -T_{11}.
\end{align}
An example where the approximation in Eq.~(\ref{eqn:1GTmT}) does not hold is the $\JP p$ scattering that produces the pentaquarks in the chiral unitary approach. In this  case
the potential  $V_{\JP p,\JP p}$ is much smaller than those for the other channels
\cite{hiddencharm}
(mostly 
$\bar D^* \Sigma_c$ and $\bar D^* \Sigma^*_c$). Let us call channel 1 the $\JP p$ and assume for simplicity that there was only another channel, $\bar D^* \Sigma_c$, (channel 2). In this case the potential matrix element $V_{12}$ is of the same order of magnitude than $V_{22}$ and $T_{12}$ is larger than $T_{11}$ \cite{hiddencharm}. Therefore $(1+GT)_{11}$ is not proportional to $T_{11}$ in the $\JP p$ scattering case.

Coming back to the meson-baryon interaction in $I=0$, $L=0$, that produces the $\Lambda(1405)$,
 one could naively think, in a first impression, that even if $V_{12}$ was not small, Eq.~(\ref{eqn:1GTmT}) should also hold next to a pole since $T_{11}$ and $T_{12}$ contain both the singularity of the pole which would factorize out. However this is true if only one pole is present, but as already mentioned above, there are two poles associated to the $\Lambda(1405)$ resonance. If we call these two poles $A$ and $B$, respectively, we would have that
\begin{align}
T_{11}\simeq \frac{g_1^A g_1^A}{\sqrt{s}-\sqrt{s_0^A}}
+\frac{g_1^B g_1^B}{\sqrt{s}-\sqrt{s_0^B}} \nn\\
T_{12}\simeq \frac{g_1^A g_2^A}{\sqrt{s}-\sqrt{s_0^A}}
+\frac{g_1^B g_2^B}{\sqrt{s}-\sqrt{s_0^B}}
\end{align}
which are indeed not proportional since the couplings of the different poles, $g_i$, to the different channels are different,
($g_1^A=2.71$, $g_1^B=2.96$, $g_2^A=3.06$ and $g_2^B=1.96$ \cite{Roca:2013cca}). 
\begin{figure}[t!]
     \centering
          \includegraphics[width=.8\linewidth]{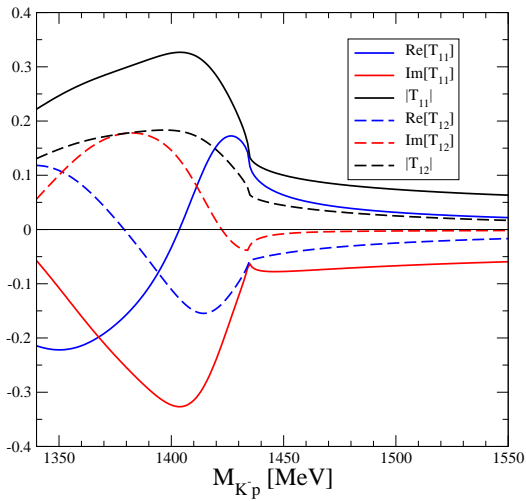}
    \caption{(Color online). Real part, imaginary part and absolute value of $t^{I=0}_{\bar K N,\bar K N}$ ($T_{11})$ and $t^{I=0}_{\bar K N,\bar \pi\Sigma}$ ($T_{12}$).}
       \label{fig:T11T12}
\end{figure}
In Fig.~\ref{fig:T11T12} we can see that $T_{11}$ and $T_{12}$ are actually very different.

\begin{figure}[t!]
     \centering
          \includegraphics[width=.8\linewidth]{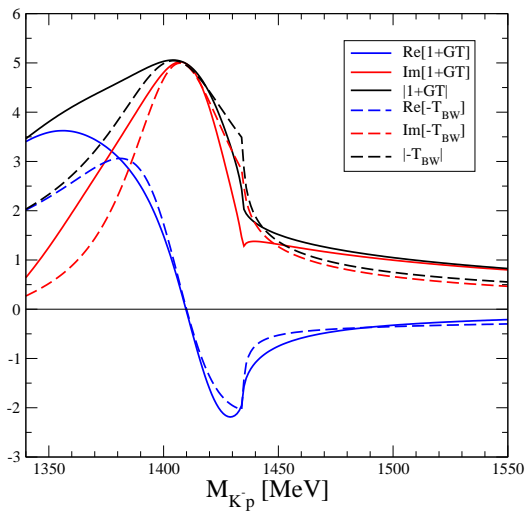}
    \caption{(Color online). Real part, imaginary part and absolute
    value of $1+G_{K^-p}\,t^{I=0}_{\bar K N,\bar K N}$ and $-T_{BW}$,
    where $T_{BW}$ is the Breit-Wigner parameterization of the
    $\Lambda(1405)$ of Eqs.~(\ref{eqn:TeqBW}) and (\ref{eqn:GBW}), the later normalized such that the absolute value agrees in the region of the $\Lambda(1405)$.}
       \label{fig:unomasGTTBW}
\end{figure}
In Fig.~\ref{fig:unomasGTTBW} we show the comparison between $(1+GT)_{11}$ and the Breit-Wigner parameterization of the $\Lambda(1405)$ of Eq.~(\ref{eqn:TeqBW}). In this figure we can see the similarity between both amplitudes, which makes the fit almost equivalent using any of them.

In conclusion, when fitting the $\Lambda(1405)$ with a 
Breit-Wigner with a Flatt\'e width,
as the experimental analysis does \cite{exp}, 
and neglecting the contact term, the result is equivalent to having considered $1+GT$ as in the chiral unitary approach. But this is true in this particular case by chance.
Note that it is a combination of the $K^-p$ tree level plus rescattering $(1+GT)$ that makes the two approaches equivalent. As mentioned, this is not trivial. Things could have also been different if, instead of having the $K^-p$ at tree level and not $\pi\Sigma$, one had the opposite situation, with $\pi\Sigma$ at tree level and not $K^-p$. We would then have something similar to the ``only $GT$'' contribution of 
Fig.~\ref{fig:res1}(a), instead of the $(1+GT)$, and the spectra of $K^-p$ would be different.

\section{Summary and conclusions}

We have addressed some interesting issues from a theoretical point of view regarding the possible pentaquark states that show up in the $\Lambda_b \to J/\psi K^- p$ decay from the LHCb experimental data \cite{exp}.
In order to model theoretically the process, we have improved over a previous work \cite{Roca:2015dva}, implementing the $K^- p$ and 
 $J/\psi p$ final state interaction and including explicitly  the relevant $\Lambda$ resonances not generated in the former rescattering. The $K^- p$ interaction generates dynamically the double $\Lambda(1405)$
pole structure implementing unitarity with $\bar K N$ and $\pi\Sigma$ coupled channels in s-wave. In the $\JP p$ final state the  possibility of the pentaquark to be dynamically generated is also allowed.
With this simplified model, as the number of $\Lambda$ resonances included with respect to the experimental analysis is concerned, but accurate enough to reproduce the experimental data, we perform several fits to the LHCb invariant mass distributions \cite{exp} for the different spin-parity possibilities for the pentaquarks.

One of the conclusions obtained is that, with the only fit to the $K^- p$ and 
 $J/\psi p$ mass distributions, the existence of the $\pa$ state cannot be undoubtedly claimed (unlike the $\pb$ state), since we get not much worse results removing this pentaquark from the fit. We have traced the origin of this similarity in the results with or without $\pa$ to the effect of a nonresonant term which provides some of the strength in the absence of the   $\pa$ resonance. 
Furthermore, we also obtain similar results for the different possibilities of the spin-parity of the pentaquarks. 
Therefore the claims regarding the existence of the $\pa$ pentaquark and the spin-parity assignments of both pentaquarks made in the experimental analysis \cite{exp} cannot be inferred just from a fit to the $K^- p$ and 
 $J/\psi p$ mass distributions. Thus, it would be most welcomed if the experimental group could singled out the observables that show unambiguously the  existence of the $\pa$ pentaquark and the spin-parity  of both of them.

On the other hand, we have widely discussed the important role played by the tree level contact elementary $\Lambda_b \to J/\psi K^- p$ production
and the $K^-p$ rescattering. The interference of these two terms gives rise to the $\Lambda(1405)$ resonance in our case, and we showed that this procedure, implementing unitarity in coupled channels, produced results in remarkable agreement with
 those of the analysis of \cite{exp}, where a $\Lambda(1405)$ resonance Breit-Wigner term (accounting for Flatt\'e effect) was introduced. We have shown that this agreement is not trivial or general, but occurs in the present case.

\section*{Acknowledgments}  

This work is partly supported
by the Spanish Ministerio de Economia y Competitividad and European
FEDER funds under contracts number FIS2011-28853-C02-01,
FIS2011-28853-C02-02, FIS2014-51948-C2-1-P, FIS2014-51948-C2-2-P, FPA2013-40483-P, FIS2014-57026-REDT and the Generalitat Valenciana in the program
Prometeo II-2014/068. 
This work is partially funded by the grants MINECO (Spain) and ERDF (EU), grant FPA2013-40483-P.

\appendix

\section{APPENDIX: Amplitudes and partial wave interferences}
\label{appendix:interf}

In this Appendix we explicitly evaluate the contribution of the different mechanisms to the total scattering amplitude  in Eq.~{\ref{eqn:dGammadM} which depend on the different partial waves and spin possibilities of the different resonances considered.

For the different spin and angular momentum we will follow the nomenclature of fig.~\ref{fig:appendix1}, {\it i.e.}, $s$, $L$ and $J^P$ are the spin, orbital angular momentum and total spin-parity respectively of the $ J/\psi p$ pair and  $L'$ and $J'^{P'}$ stand for the $K^-$ orbital angular momentum and total spin-parity of the $p K^-$ system.

\begin{figure}[tbp]
     \centering
              \includegraphics[width=.7\linewidth]{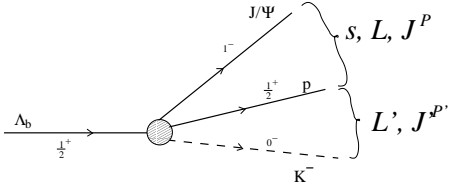}
    \caption{Spin and angular momentum nomenclature for the general $\Lambda_b \to J/\psi K^- p$ process}
    \label{fig:appendix1}
\end{figure}

If we assume that the dominant contribution to a given partial wave is given by the smallest allowed orbital angular momentum, we can  write the relevant terms of the $\Lambda_b \to J/\psi K^- p$ amplitude of interest in the present work as
\ba
T= a S_{\frac{1}{2}^-} + b P_{\frac{3}{2}^-} + c P_{\frac{1}{2}^-} + e D_{\frac{3}{2}^-} + f D_{\frac{5}{2}^+},
\label{eq:Tstructure}
\ea
where the nomenclature of the different terms stands for $L'_{J^P}$
and correspond to the quantum numbers shown in table~\ref{tab:appendix1}.
(The amplitude of  Eq.~(\ref{eq:Tstructure}) contains only the relevant terms for the quantum numbers of the pentaquarks and $\Lambda$ resonances considered in the present work). In the last line of the table the different $\Lambda$ resonances contributing to a given partial wave are shown.
(Note that in the $P_{\frac{3}{2}^-}$ and  $P_{\frac{1}{2}^-}$ columns, the 
$J'^{P'}$ can also be $\frac{3}{2}^+$ but we do not need it since we do not consider any $\Lambda$ resonances with that $J'^{P'}$).
\begin{table}[h]
\caption{Quantum numbers of the different contributions to the  $\Lambda_b \to J/\psi K^- p$  amplitude with the lowest possible value of orbital angular momentum for a given value of $J^P$ and the $J'^{P'}$ used in the present work.}
\begin{center}
\begin{tabular}{|c |c| c| c| c| c|} 
\hline    
      &   $S_{\frac{1}{2}^-}$ & $P_{\frac{3}{2}^-}$ & $P_{\frac{1}{2}^-}$&  $D_{\frac{3}{2}^-}$  & $D_{\frac{5}{2}^+}$ \\ \hline
$s$   &  $\frac{1}{2}$     &  $\frac{3}{2}$    &  $\frac{1}{2}$ &  $\frac{3}{2}$    &  $\frac{3}{2}$\\  \hline
$L$   &  $0$     &  $0$    &  $0$ &  $0$    &  $1$\\  \hline
$J^P$ &  $\frac{1}{2}^-$     &  $\frac{3}{2}^-$    &  $\frac{1}{2}^-$ &  $\frac{3}{2}^-$    &  $\frac{5}{2}^+$\\  \hline
$L'$  &  $0$     &  $1$    &  $1$ &  $2$    &  $2$\\  \hline
$J'^{P'}$&  $\frac{1}{2}^-$     &  $\frac{1}{2}^+$    &  $\frac{1}{2}^+$ &  $\frac{3}{2}^-$    &  $\frac{5}{2}^-$\\  \hline
$\Lambda$&  \parbox[t]{1.1cm}{$\Lambda(1405)$\\$\Lambda(1800)$}     &  \parbox[t]{1.1cm}{$\Lambda(1600)$\\$\Lambda(1810)$}    & \parbox[t]{1.1cm}{$\Lambda(1600)$\\$\Lambda(1810)$}    & \parbox[t]{1.1cm}{$\Lambda(1520)$\\$\Lambda(1690)$}    &  \\ \hline
 \end{tabular}
\end{center}
\label{tab:appendix1}
\end{table}

The term $S_{\frac{1}{2}^-}$ term is given by $S_{\frac{1}{2}^-}=\vec \sigma\cdot\vec \epsilon$ where $\vec\epsilon$ stands for the $\JP$ polarization vector. This operator projects over $J/\psi p$ and $K^-$ s-waves \cite{Garzon:2012np}.
The structures  
        $P_{\frac{1}{2}^-}$ and $P_{\frac{3}{2}^-}$ are obtained in
 ref.~ \cite{Lu:2016roh} by imposing orthonormality to  $S_{\frac{1}{2}^-}$ and are given by \footnote{Note the different nomenclature with respect to reference \cite{Lu:2016roh}:
  $P_{\frac{3}{2}^-}$ corresponds to $S_{3/2}$ of \cite{Lu:2016roh} and 
 $P_{\frac{1}{2}^-}$ corresponds to $S_{1/2}$ of \cite{Lu:2016roh}}:
 \ba
 P_{\frac{3}{2}^-}&=&  \langle m_p \mid  k_j\, \epsilon_j + \frac{i}{2} \,\epsilon_{ijl} \,\sigma_l \, k_i \epsilon_j \mid m_{\Lambda_b}\rangle, \nn \\
P_{\frac{1}{2}^-}&=&  \langle m_p \mid  k_j\, \epsilon_j -  i \,\epsilon_{ijl} \,\sigma_l \, k_i \epsilon_j \mid m_{\Lambda_b}\rangle,  \ea
where $k$ is the $K^-$ momentum.
Note that the $\Lambda(1600)$ and $\Lambda(1810)$ contribute both to 
$P_{\frac{3}{2}^-}$ and $P_{\frac{1}{2}^-}$. Since the first $\Lambda_b \JP \Lambda$ vertex is s-wave and the $\Lambda p K^-$ vertex  is p-wave, the actual structure for the $\Lambda(1600)$ and $\Lambda(1810)$ cases is
 $ \langle m_p | \vec\sigma\cdot\vec k\, \vec\sigma\cdot\vec\epsilon| m_{\Lambda_b}\rangle$ which in terms of 
$P_{\frac{3}{2}^-}$ and $P_{\frac{1}{2}^-}$ reads
\be
\langle m_p | \vec\sigma\cdot\vec k\, \vec\sigma\cdot\vec\epsilon| m_{\Lambda_b}\rangle= \frac{4}{3}P_{\frac{3}{2}^-}-\frac{1}{3}P_{\frac{1}{2}^-}.
\ee
Although not needed in the present work, the $J'^{P'}=\frac{3}{2}^+$ combination is given by
\be
\langle m_p | \vec S\cdot\vec k\, \vec S^\dagger\cdot\vec\epsilon| m_{\Lambda_b}\rangle= \frac{2}{9}P_{\frac{3}{2}^-}+\frac{4}{9}P_{\frac{1}{2}^-},
\ee
with $\vec S^\dagger$ the spin $1/2$ to $3/2$ transition operator.

On the other hand, $D_{\frac{3}{2}^-}$ is given by
\be
D_{\frac{3}{2}^-}=\langle m_p \mid  
(k_i k_j -\frac{1}{3}\vec k^2\delta_{ij})\sigma_i\epsilon_j
 \mid m_{\Lambda_b}\rangle
\label{eq:D321}
\ee
since it accounts for the $K^-$ in d-wave and $J/\psi p$ in s-wave. 
Finally, the expression for the $D_{\frac{5}{2}^+}$ is
\be
D_{\frac{5}{2}^+}=\langle m_p \mid  
i(\vec\sigma\times\vec\epsilon)_i p_j(k_i k_j -\frac{1}{3}\vec k^2\delta_{ij})
 \mid m_{\Lambda_b}\rangle.
\label{eq:D52p}
\ee
where $p$ is the $\JP$ momentum.
Note that, while the $(k_i k_j -\vec k^2\delta_{ij}/3)$ term in Eq.~(\ref{eq:D52p}) is purely d-wave, the $i(\vec\sigma\times\vec\epsilon)_i p_j$ term gives contribution to both 
$\frac{3}{2}^+$ and $\frac{5}{2}^+$ for the $J/\psi p$ pair. However the $\frac{3}{2}^+$ is also attainable with $L'=1$, which is reasonably the dominant contribution for  $\frac{3}{2}^+$, while for  $\frac{5}{2}^+$  $L'=2$ is the lowest  possible $L'$. This is the reason why we consider Eq.(\ref{eq:D52p}) to account  for the $J^P=\frac{5}{2}^+$ case.

The sum over initial and final spins of the total amplitude in Eq.~(\ref{eq:Tstructure}) gives
\ba
\sum |T|^2&=&3 |a|^2+|b|^2 \frac{3}{2}\vec k^2+|c|^2 3\vec k^2
+|e|^2 \frac{2}{3}\vec k^4 \nn\\
&+& |f|^2 \frac{2}{3}\vec k^2
\left[(\vec p\cdot\vec k)^2+\frac{1}{3}\vec k^2\vec p\,^2\right]\nn\\
&-&\frac{4}{3} \vec k^2\vec p\cdot\vec k\left[ \textrm{Re}(b f^*)
-2\,\textrm{Re}(c f^*)\right],
\ea
where $|\vec k|=\lambda^{1/2}(M_{K^-p}^2,m_p^2,m_K^2)/(2M_{K^-p})$ is the kaon momentum in the $K^-p$ rest frame and 
\be\vec p\cdot\vec k=\frac{1}{2}\left(M_{\JP\,p}^2+2 \tilde P^0 \tilde k^0
-M_{\Lambda_b}^2-m_K^2\right)
\ee
 where $\tilde P^0$,  $\tilde k^0$, are the energies of the $\Lambda_b$, $K^-$, in the $K^-p$ rest frame:
\ba
\tilde P^0&=&\sqrt{M_{\Lambda_b}^2+\tilde p^2}\nn\\
\tilde k^0&=&\frac{M_{K^-p}^2+m_K^2-m_p^2}{2M_{K^-p}},
\ea
with
\be
\tilde p=\frac{\lambda^{1/2}(M_{\Lambda_b}^2,M_{K^-p}^2,M_{\JP}^2)}
{2M_{K^-p}}.
\ee

To the coefficients $a$, $b$, $c$, $e$ and $f$ contribute the different mechanisms explained in formalism section according to their respective quantum numbers:

\begin{align}
a=&\alpha_1 \left(1+G_{K^-p}(M_{K^-p})\,t^{I=0}_{\bar K N,\bar K N}(M_{K^-p})\right)\nn\\
+&\delta_{J^P_B,\frac{1}{2}^-}\alpha_2\,G_{\JP p}\,\frac{g^2_{\JP\,p}}{M_{\JP\,p}-m_{P_c(4450)}+i \frac{\Gamma_{P_c(4450)}}{2}}\nn\\
+&\delta_{J^P_A,\frac{1}{2}^-}\alpha_3\,G_{\JP p}\,\frac{g^2_{\JP\,p}}{M_{\JP\,p}-m_{P_c(4380)}+i \frac{\Gamma_{P_c(4380)}}{2}}\nn\\
+&\alpha_4\frac{1}{ {M_{K^-p}-m_{\Lambda(1800)}+i\frac{\Gamma_{\Lambda(1800)}}{2}}},
\label{eq:coeffa}
\end{align}

\begin{align}
b&=\frac{4}{3} \alpha_5 \frac{1}{ {M_{K^-p}-m_{\Lambda(1600)}+i\frac{\Gamma_{\Lambda(1600)}}{2}}}\nn\\
&+\frac{4}{3}\alpha_6 \frac{1}{    {M_{K^-p}-m_{\Lambda(1810)}+i\frac{\Gamma_{\Lambda(1810)}}{2}}}
\nn\\
&+
C_{3/2}\bigg[1+\nn\\
&+\delta_{J^P_B,\frac{3}{2}^-}\alpha_2\,G_{\JP p}\,\frac{g^2_{\JP\,p}}{M_{\JP\,p}-m_{P_c(4450)}+i \frac{\Gamma_{P_c(4450)}}{2}}\nn\\
&+\delta_{J^P_A,\frac{3}{2}^-}\alpha_3\,G_{\JP p}\,\frac{g^2_{\JP\,p}}{M_{\JP\,p}-m_{P_c(4380)}+i \frac{\Gamma_{P_c(4380)}}{2}}\bigg]
\label{eq:coeffb}
\end{align}

\begin{align}
c=&-\frac{1}{3} \alpha_5 \frac{1}{ {M_{K^-p}-m_{\Lambda(1600)}+i\frac{\Gamma_{\Lambda(1600)}}{2}}}\nn\\
&-\frac{1}{3}\alpha_6 \frac{1}{    {M_{K^-p}-m_{\Lambda(1810)}+i\frac{\Gamma_{\Lambda(1810)}}{2}}}
\nn\\
&+
C_{1/2}\bigg[1+\nn\\
&+\delta_{J^P_B,\frac{1}{2}^-}\alpha_2\,G_{\JP p}\,\frac{g^2_{\JP\,p}}{M_{\JP\,p}-m_{P_c(4450)}+i \frac{\Gamma_{P_c(4450)}}{2}}\nn\\
&+\delta_{J^P_A,\frac{1}{2}^-}\alpha_3\,G_{\JP p}\,\frac{g^2_{\JP\,p}}{M_{\JP\,p}-m_{P_c(4380)}+i \frac{\Gamma_{P_c(4380)}}{2}}\bigg]
\label{eq:coeffc}
\end{align}

\begin{align}
e=&\alpha_7 \frac{1}{ {M_{K^-p}-m_{\Lambda(1520)}+i\frac{\Gamma_{\Lambda(1520)}}{2}}} \nn\\
+
&\alpha_8 \frac{1}{ {M_{K^-p}-m_{\Lambda(1690)}+i\frac{\Gamma_{\Lambda(1690)}}{2}}}
\label{eq:coeffe}
\end{align}

\begin{align}
f=&
C_{5/2}\bigg[1+\nn\\
&+\delta_{J^P_B,\frac{5}{2}^+}\alpha_9
\,\frac{1}{M_{\JP\,p}-m_{P_c(4450)}+i \frac{\Gamma_{P_c(4450)}}{2}}\nn\\
&+\delta_{J^P_A,\frac{5}{2}^+}\alpha_{10}
\,\frac{1}{M_{\JP\,p}-m_{P_c(4380)}+i \frac{\Gamma_{P_c(4380)}}{2}}\bigg].
\label{eq:coefff}
\end{align}

In the previous equations $J^P_A$ and $J^P_B$ stand for the spin-parity of the $P_c(4380)$ and $P_c(4450)$ respectively.
The Kronecker deltas are introduced to account for the different possibilities of quantum numbers of the two pentaquarks which, according to their $J^P$, contribute to the corresponding partial wave amplitude.
For instance, if one wanted to consider the $(J^P_A,J^P_B)=(1/2^-,5/2^+)$ case
then the $\pa$ propagator would contribute to the $a$ and $c$ coefficient, while the $\pa$ propagator would contribute only to the $f$ coefficient. 

Note the extra inclusion of contact terms (the 1 addends) in $b$, $c$ and $f$ coefficients to account for possible contact terms with those quantum numbers. 

The $\alpha_i$ and $C_i$ coefficients are complex in general. Therefore, taking into account that there are some unobservable arbitrary global phases, there are 19 free parameters to be fitted in the general case.

\end{document}